\documentclass[printer]{aa}
\usepackage{graphicx,natbib,amsmath}
\usepackage{epic,eepic}

\begin{document}
 
\def\valid{}   

\font\caps=cmcsc10                  
\font\dunh=cmdunh10  at 12.0 true pt 
\font\dunhs=cmdunh10 
\font\vbold=cmbx10 scaled \magstep1 
\font\sevenbf=cmbx7
\font\sevenit=cmti7
\font\Kapi=cmr17

\def\MEV{DOME}
\def\RTE{equation of radiative transfer}
\def\etal{{et al}}
\def\HW{H\&W}
\def\OK{O\&K}
\def\ok{O\&K}
\def\RH{R\&H}

\def\ibmrs{\hbox{\tt RS/6000}}
\def\hp{\hbox{\tt HP~9000}}
\def\dec{\hbox{\tt DEC~5000}}
\def\axp{\hbox{\tt AXP}}
\def\ibmmf{\hbox{\tt IBM~3090}}
\def\ibmpc{\hbox{\tt 486DX}}
\def\cray{\hbox{\tt Cray 2}}
\def\ymp{\hbox{\tt YMP}}
\def\nec{\hbox{\tt NEC}}

\def\g{\gamma}
\def\b{\beta}
\def\m{\mu}
\def\e{\epsilon}
\def\n{\nu}
\def\l{\lambda}
\def\L{\Lambda}
\def\K{{\rm K}}
\def\logg{{\log(g)}}
\def\t{\tau}
\def\pder#1#2{{\partial #1 \over \partial #2}}
\def\div#1#2{{#1\over #2}}
\def\rout{\ifmmode{r_{\rm out}}\else\hbox{$r_{\rm out}$}\fi}
\def\tmax{\ifmmode{\tau_{\rm max}}\else\hbox{$\tau_{\rm max}$}\fi}
\def\tstd{\ifmmode{\tau_{\rm std}}\else\hbox{$\tau_{\rm std}$}\fi}
\def\vmax{\ifmmode{v_{\rm max}}\else\hbox{$v_{\rm max}$}\fi}
\def\muE{\ifmmode{\mu_{\rm E}}\else\hbox{$\mu_{\rm E}$}\fi} 
\def\pE{\ifmmode{p_{\rm E}}\else\hbox{$p_{\rm E}$}\fi} 
\def\bmax{\ifmmode{\b_{\rm max}}\else\hbox{$\b_{\rm max}$}\fi}
\def\kms{\hbox{$\,$km$\,$s$^{-1}$}}
\def\ergs{\hbox{$\,$erg$\,$s$^{-1}$}}
\def\kpc{\hbox{$\,$kpc} }
\def\ang{\hbox{\AA}}
\def\Msun{\hbox{$\,$M$_\odot$} }
\def\Lsun{\hbox{$\,$L$_\odot$} }
\def\Teff{\hbox{$\,T_{\rm eff}$} }
\def\alog#1{\times 10^{#1}}
\def\rin{\hbox{$r_{\rm in}$} }
\def\rout{\hbox{$r_{\rm out}$} }

\def\lstar{\ifmmode{\Lambda^*}\else\hbox{$\Lambda^*$}\fi} 
\def\Lstar{\ifmmode{\Lambda^*}\else\hbox{$\Lambda^*$}\fi} 
\def\Rop{\ifmmode{[R_{ij}]}\else\hbox{$[R_{ij}]$}\fi}
\def\Rij{\Rop}
\def\Rji{\ifmmode{[R_{ji}]}\else\hbox{$[R_{ji}]$}\fi}
\def\Rstar{\ifmmode{[R_{ij}^*]}\else\hbox{$[R_{ij}^*]$}\fi}
\def\Rijstar{\Rstar}
\def\Rjistar{\ifmmode{[R_{ji}^*]}\else\hbox{$[R_{ji}^*]$}\fi}
\def\DRji{\ifmmode{[\Delta R_{ji}]}\else\hbox{$[\Delta R_{ji}]$}\fi}
\def\DRij{\ifmmode{[\Delta R_{ij}]}\else\hbox{$[\Delta R_{ij}]$}\fi}

\def\Jb{{\bar J}}
\def\Jbar{{\bar J}}
\def\Jnew{{\bar J_{\rm new}}}
\def\Jold{{\bar J_{\rm old}}}
\def\Jfs{{\bar J_{\rm fs}}}
\def\Snew{{S_{\rm new}}}
\def\Sold{{S_{\rm old}}}
\def\Amat{\mat{A}}             

\def\ns{\ifmmode{N_{\rm s}}          
        \else\hbox{$N_{\rm s}$}\fi}
\def\ion#1{\hbox{ #1}}

\def\peq{\mathbin{\hbox{$+$}\hbox{$=$}}}

\def\mat#1{{\bf #1}}     
\def\vek#1{{#1}}         

\newcount\eqcount
\eqcount=0
\def
  \nummer{
    \global\advance\eqcount by 1
    (\the\eqcount)
  }

\def
  \numadv{
    \global\advance\eqcount by 1
  }

\def
   \numout#1{
     (\the\eqcount #1)
  }

\def\ivek#1#2{\ifmmode{\vek{I}^{#1}_{#2}}
        \else\hbox{$\vek{I}^{#1}_{#2}$}\fi}

\def\ip#1{\ivek{+}{#1}}      
\def\im#1{\ivek{-}{#1}}

\def\tmat#1#2{\ifmmode{\mat{t}^{#1}_{#2}}
        \else\hbox{$\mat{t}^{#1}_{#2}$}\fi}
\def\rmat#1#2{\ifmmode{\mat{r}^{#1}_{#2}}
        \else\hbox{$\mat{r}^{#1}_{#2}$}\fi}
\def\bvek#1#2{\ifmmode{\beta^{#1}_{#2}}
        \else\hbox{$\beta^{#1}_{#2}$}\fi}

\def\tpi#1{\tmat{+}{#1}}
\def\tmi#1{\tmat{-}{#1}}
\def\rmi#1{\rmat{-}{#1}}
\def\rpi#1{\rmat{+}{#1}}
\def\bpi#1{\bvek{+}{#1}}
\def\bmi#1{\bvek{-}{#1}}

\def\tp{\tmat{+}{}}          
\def\tm{\tmat{-}{}}          
\def\rmm{\rmat{-}{}}         
\def\rp{\rmat{+}{}}          
\def\bp{\bvek{+}{}}          
\def\bm{\bvek{-}{}}          
\def\tpm{\tmat{\pm}{}}       
\def\rpm{\rmat{\pm}{}}       
\def\bpm{\bvek{\pm}{}}

\def\lp{\ifmmode{\lambda^+_\tau}           
        \else\hbox{$\lambda^+_\tau$}\fi}
\def\lm{\ifmmode\lambda^-_\tau             
        \else\hbox{$\lambda^-_\tau$}\fi}

\def\phxO{{\tt PHOENIX/1D}}
\def\phxT{{\tt PHOENIX/3D}}
\def\phx{{\tt PHOENIX}}

\baselineskip=12pt

\title{A 3D radiative transfer framework: XI. multi-level NLTE}

\titlerunning{3D radiative transfer framework XI}
\authorrunning{Hauschildt and Baron}
\author{Peter H. Hauschildt\inst{1} and E.~Baron\inst{1,2}}

\institute{
Hamburger Sternwarte, Gojenbergsweg 112, 21029 Hamburg, Germany;
yeti@hs.uni-hamburg.de 
\and
Homer L.~Dodge Dept.~of Physics and Astronomy, University of
Oklahoma, 440 W.  Brooks, Rm 100, Norman, OK 73019 USA;
baron@ou.edu
}

\date{Received date \ Accepted date}

\abstract
{Multi-level non-local thermodynamic equilibrium (NLTE) radiation transfer calculations have become standard throughout
  the stellar atmospheres community and are applied to all types of
  stars as well as dynamical systems such as novae and
  supernovae. Nevertheless even today spherically symmetric 1D calculations
with full physics are computationally intensive. We show that full
physics NLTE calculations can be done with fully 3 dimensional (3D)
radiative transfer. }
{
With modern computational techniques and current massive parallel
computational resources, full detailed solution of the multi-level NLTE problem
coupled to the solution of the radiative transfer scattering problem
can be solved without sacrificing the micro physics description. 
}
{
We extend the use of a rate operator developed to solve the coupled
NLTE problem in spherically symmetric 1D systems.
In order to spread
memory among processors  
we
have implemented the NLTE/3D module with a hierarchical domain decomposition
method that distributes the NLTE levels, radiative rates, and rate operator data
over a group of processes so that each process only holds the data for a
fraction of the voxels. Each process in a group holds all the relevant data to
participate in the solution of the 3DRT problem  so
that the 3DRT solution is parallelized within a domain decomposition group.
}
{
We solve a spherically symmetric system  in 3D spherical coordinates in order to directly compare
our well-tested 1D code to the 3D case. We compare three levels of
tests: a) a simple H+He test calculation, b) H+He+CNO+Mg, c)
H+He+Fe. The last test is computationally large and shows that
realistic astrophysical problems are solvable now, but they do require
significant computational resources.
}
{With presently available computational resources it is possible to solve the
full 3D multi-level problem with the same detailed 
micro-physics as included in 1D modeling.}


\maketitle

\section{Introduction}

In a series of papers \citet*[][hereafter: Papers I--X]{3drt_paper1,
3drt_paper2, 3drt_paper3, 3drt_paper4,
3drt_paper5,3drt_paper7,3drt_paper8,3drt_paper9,3drt_paper10}, we have
described a framework for the solution of the radiative transfer equation in 3D
systems (3DRT), including a detailed treatment of scattering in continua and
lines with a non-local operator splitting method. In \citet{3drt_paper6} we
described tests of the 3D mode of the \phx\ model atmosphere code package.

Here, we describe the implementation and the results of detailed
multi-level non-local thermodynamic equilibrium (NLTE) \phxT\ calculations and
compare the results to equivalent 1D calculations with \phxO\ models. We will
first describe the method we have implemented and discuss differences to the 1D
version, then we will show and discuss the results of simple test calculations.

As 3D hydrodynamical calculations become more common, detailed
radiative transfer effects due to the 3D structure will be needed in
order to directly compare the predictions of hydrodynamic results with
observations. 3D effects due to convective structure are known to be
important in the sun and other stars
\citep{hayek11,AGS02,AspIII00,ANTS00,ANTS99}. It is also known that
NLTE effects can play an important role \citep{bergemann12}. This is
also the case for brown dwarfs, irradiated planets, and circumstellar
disks \citep{huegelmeyer09,wawrzyn09,witte11}. 3D radiative transfer
effects play a role in interpreting the spectra of active stars
\citep{berkner13}. In addition, 3D radiative transfer effects are
important in the binary environment of Type Ia supernovae
\citep{kasen_hole04,thomas02,kasen01el03} and in the disks of AGN.
 Here, we present results for the 3D spherical coordinate
system mode of \phxT. However, the method is coordinate system indenpendent.

\section{Method}

In the following discussion we use  notation of Papers I -- X.  The basic
framework and the methods used for the formal solution and the solution of the
scattering problem via  non-local operator splitting are discussed in detail in
these papers and will not be repeated here. The algorithm and implementation of
the 3D NLTE module (NLTE/3D) follows our 1D method \citep[]{casspap}, however,
we will include an updated detailed description from \cite{casspap} here for
convenience and easier discussion.

\section{NLTE/3D implementation}

\subsection{The rate equations}

For each voxel, the NLTE rate equations have the form \citep[e.g.,][]{mihalas78}
\begin{multline}
    \sum_{j<i} n_j \left(R_{ji}+C_{ji}\right) \\
{   -n_i\left\{\sum_{j<i} \left(\div{n_j}{n_i}\right)^{*}
                \left(R_{ij}+C_{ji}\right)
             +\sum_{j>i}
                \left(R_{ij}+C_{ij}\right)\right\} } \\
{    +\sum_{j>i} n_j
     \left(\div{n_i}{n_j}\right)^{*}\left(R_{ji}+C_{ij}\right)   = 0.}
\label{REQ}
\end{multline}
In Eq.~\ref{REQ}, $n_i$ is 
the actual, NLTE population density of a level $i$
and the symbol $n_i^{*}$ denotes the so-called LTE population density
of the level $i$, which is given in the ``Menzel definition'' \citep{MC37,mihalas78} by 
\begin{equation}
     {n_i^{*}}
      = \div{g_i}{g_\kappa} {n_\kappa}{n_e }
       {2 h^3  \over (2 \pi m kT)^{3/2} }
       \exp\left(\div{\chi_{\text{ion}}-\chi_i}{kT}\right).\label{bi-def}
\end{equation}
Here $n_\kappa$ denotes the \textit{actual}, i.e., NLTE,
population density of the ground
state of the next higher ionization stage of the same element;
${g_i}$ and ${g_\kappa}$ are
the statistical weights of the levels $i$ and $\kappa$, respectively.
In Eq.~\ref{bi-def}, $\chi_i$ is the excitation energy of the level $i$ and
$\chi_{\text{ion}}$ denotes the ionization energy from the ground state to 
the corresponding ground state of the next higher ionization stage.
The actual, NLTE electron density is given by $n_e$. The system of
rate equations is closed by the conservation equations for the nuclei and the 
charge conservation equation \citep[]{mihalas78}.

The rates for radiative and collisional transitions between two levels
$i$ and $j$ (including transitions from and to the continuum, see below)
are given by
$R_{ij}$ and $C_{ij}$, respectively. 
We will use $J_\lambda$ rather than the more conventional $J_\nu$, therefore, the upward (absorption) radiative rates $R_{ij}$ 
($i<j$) 
are given by
\[
   R_{ij} = 
    {4\pi\over hc}
     \int_0^\infty \alpha_{ij}(\lambda)
J_\lambda(\lambda)\,\lambda d\lambda, 
\]
whereas the downward (emission) radiative rates $R_{ji}$ ($i<j$) are 
given by
\[
   R_{ji}= 
    {4\pi\over hc}
     \int_0^\infty \alpha_{ji}(\lambda)
     \left( {2hc^2\over\lambda^5} +J_\lambda(\lambda)\right) 
     \exp\left(-{hc\over k\lambda T}\right)\,\lambda d\lambda.
\]
Here, $J_\lambda$ is the mean intensity, $T$ the electron temperature, $h$ and
$c$ and Planck's constant and the speed of light, respectively. 

We also follow the convention of \citet{mihalas78} that since
\[
n_l^*C_{lu} = n_u^*C_{ul}
\]
then
\[
n_uC_{ul} = n_u \left(\frac{n_l}{n_u}\right)^*C_{lu}
\]
and therefore only \textit{upward} collision rates $C_{lu}$ appear in
the rate equations (Eq.~\ref{REQ}).

The 
cross section $\alpha_{ij}(\lambda)$ of the transition $i\to j$ at the
wavelength $\lambda$  for bound-bound transitions is given by
\[
   \alpha_{ij}(\lambda)=\hat\sigma_{ij} \varphi_\lambda(\lambda)
  = 
   \div{hc}{4\pi}\div{\lambda_{ij}}{c}B_{ij} \varphi_\lambda(\lambda),
\]
and
\[
   \alpha_{ji}(\lambda)=\hat\sigma_{ij} \phi_\lambda(\lambda)
  = 
   \div{hc}{4\pi}\div{\lambda_{ij}}{c}B_{ij} \phi_\lambda(\lambda),
\]
where $\lambda_{ij}$ and $B_{ij}$ are the rest wavelength and
the Einstein coefficient for absorption  of the transition 
$i\to j$,
respectively; 
$\varphi_\lambda(\lambda)$ is the normalized absorption profile, whereas
$\phi_\lambda(\lambda)$ is the normalized emission profile. In the special case
of complete redistribution (CRD), we have 
$\varphi_\lambda(\lambda)= \phi_\lambda(\lambda)$ and, therefore, 
$\alpha_{ij}(\lambda)=\alpha_{ji}(\lambda)$.

The emission coefficient $\eta_{ij}(\lambda)$ 
for a bound-bound transition is then given by
\[ 
   \eta_{ij}(\lambda)= {2hc^2\over\lambda^5}\div{g_i}{g_j} 
     \alpha_{ji}(\lambda) n_j 
\]
and the absorption coefficient is 
\[
  \kappa_{ij}(\lambda)=\alpha_{ij}(\lambda) n_i 
   - \alpha_{ji}(\lambda)\div{g_i}{g_j} n_j 
\]
For photo ionization and photo recombination transitions, the 
corresponding coefficients are 
\begin{equation}
  \eta_{i\kappa}(\lambda)={2hc^2\over\lambda^5} 
\alpha_{i\kappa}(\lambda)
      n_\kappa^{*} \exp\left(-{hc\over k\lambda T}\right)\label{eq:etadef}
\end{equation}
and
\[
  \kappa_{i\kappa}(\lambda)=\left[n_i
    -n_\kappa^{*} \exp\left(-{hc\over k\lambda T}\right)\right] 
      \alpha_{i\kappa}(\lambda)
\]

The total absorption $\chi(\lambda)$ and emission $\eta(\lambda)$ 
coefficients are obtained by summing up the contributions of 
all transitions, i.e.
\[
   \eta(\lambda) = \sum_{i<j} \eta_{ij}(\lambda) + \tilde\eta(\lambda)
\]
and
\begin{equation}
  \chi(\lambda) = \sum_{i<j} \kappa_{ij}(\lambda) + 
\tilde\kappa(\lambda)
     +\tilde\sigma(\lambda), \label{chidef}
\end{equation}
where $\tilde\eta(\lambda)$, $\tilde\kappa(\lambda)$ and 
$\tilde\sigma(\lambda)$
summarize background emissivities, absorption and scattering coefficients, 
respectively.

\subsection{3.\ The rate operator}

In this section, we rewrite the rate equations in the form of 
an `operator equation'. This equation is then used to introduce an 
`approximate rate operator' in analogy to the approximate 
$\L$-operator which can then be 
iteratively solved by an operator splitting method, following 
the ideas of \citet{rh91}.

We first introduce the `rate operator' $[R_{ij}]$ for upward transitions
in analogy to the $\Lambda$-operator. $[R_{ij}]$ is defined so that
\[
 R_{ij} = [R_{ij}][n].
\]
Here, $[n]$ denotes the `population density operator', which 
can be considered as the vector of the population densities of all 
levels at all points in the medium under consideration. 
The radiative rates are (linear) functions of the mean intensity $J$, 
which is given by
$J(\lambda) = \Lambda(\lambda) S(\lambda)$, where 
$S=\eta(\lambda)/\chi(\lambda)$ 
is the source function.
Using the $\Lambda$-operator, we can write $[R_{ij}][n]$ as:
\[
    [R_{ij}][n]={4\pi\over hc}
       \int \alpha_{ij}(\lambda)\Lambda(\l) S(\lambda) \,\lambda d\lambda.
\]
Following \citet{rh91}, we rewrite the $\Lambda$-operator as 
$$
      \Lambda(\lambda)=\Psi(\lambda) [1/\chi(\lambda)],
$$ 
where we have introduced the $\Psi$-operator \citep[see,][]{rh91}  and 
$[1/\chi(\lambda)]$ is
the diagonal operator of multiplying by $1/\chi(\lambda)$. 
Using the $\Psi$-operator, we can write $[R_{ij}]$ as
\[
   [R_{ij}][n] = {4\pi\over hc}
     \int \alpha_{ij}(\lambda)\Psi(\lambda)\eta(\lambda)\,\lambda d\lambda
\]
where $\eta(\lambda)$ is a function of the population densities and the
background emissivities. Using 
Eq.~\ref{eq:etadef} we can write $\eta(\l)$ as
\[ 
   \eta(\lambda) = \sum_{i<j} \eta_{ij}(\lambda) + \tilde\eta(\lambda)
    \equiv [E(\lambda)][n], 
\] 
where we have defined the linear and diagonal operator $[E(\l)]$.
We write the total
contribution of a particular level $k$ to the emissivity as
\begin{multline}
 \eta_k(\l) = {2hc^2\over\lambda^5} \Bigg\{
	\sum_{l} \div{g_l}{g_k} \alpha_{kl}(\lambda) \\
  + \sum_{l} \alpha_{kl}(\lambda) \exp\left(-{hc\over k\lambda T}\right)
	  \div{g_l}{g_k} \\
	  \times {2 h^3 n_e  \over (2 \pi m)^{3/2} (kT)^{3/2}}
	   \exp\left(-\div{\chi_l-\chi_k}{kT}\right)
      \Bigg\} n_k \\
 \equiv E_k(\lambda) n_k
\end{multline}
where the first sum is the contribution of the level $k$ to all bound-bound
transitions and the second sum is the contribution to all bound-free transitions. 
Therefore, $[E(\lambda)][n]$ has the form
\[
      [E(\lambda)][n] = \sum_k E_k(\lambda)n_k + \tilde \eta(\lambda).
\]

Using the $[E(\l)]$-operator, we write $\Rop [n]$ in the form
\[
     \Rop [n] = {4\pi\over hc}
     \left[ \int_0^\infty \alpha_{ij}(\lambda)
	 \Psi(\lambda) E(\lambda) \,\lambda d\lambda \right] [n]. 
\]
The corresponding expression for the emission rate-operator $[R_{ji}]$ is given by
\begin{multline}
  \Rji [n] = \\ 
  {4\pi\over hc}
     \int_0^\infty \alpha_{ji}(\lambda)
     \left\{ {2hc^2\over\lambda^5} +\Psi(\lambda) [E(\lambda)][n]\right\} 
     \exp\left(-{hc\over k\lambda T}\right)\,\lambda d\lambda   
\end{multline}
Using the rate operator, we can write the rate equations in the form
\begin{multline}
    \sum_{j<i} n_j \left([R_{ji}][n]+C_{ji}\right)\\
   -n_i\left\{\sum_{j<i} \left(\div{n_j^{*}}{n_i^{*}}\right)
		\left([R_{ij}][n]+C_{ji}\right)
	     +\sum_{j>i} 
		 \left([R_{ij}][n]+C_{ij}\right)\right\}      \\
   +\sum_{j>i} n_j 
     \left(\div{n_i^{*}}{n_j^{*}}\right)\left([R_{ji}][n]+C_{ij}\right)
     = 0.
\end{multline}
This form shows, explicitly, the non-linearity of the rate equations with 
respect to the
population densities. Note, that, in addition, the rate equations are
non-linear with respect to the electron density via the collisional 
rates and the 
charge conservation constraint condition. 

As in the case of the two-level atom, a simple $\L$-iteration scheme
will converge much too slowly to be useful for most
cases of practical
interest. Therefore, we split the rate operator, in analogy to the 
splitting of the 
$\Lambda$-operator, by $\Rij=\Rijstar + (\Rij-\Rijstar)\equiv 
\Rijstar+\DRij$,   where $\Rijstar$ is the
``approximate rate-operator''. We then rewrite the rate $R_{ij}$ as 
$$
     R_{ij} = \Rijstar [n_{\text{new}}] + \DRij [n_{\text{old}}].     \label{Rop}
$$
Analogously, we can make the same definitions  for the downward
radiative rates. In Eq.~\ref{Rop},  
$[n_{\text{old}}]$
denotes the current (old) population densities, whereas $[n_{\text{new}}]$ 
are the updated (new) population densities to be calculated. The $\Rijstar$ and
$\Rjistar$ are linear functions of the population density operator $[n_k]$ of
any level $k$, due to the linearity of $\eta$ and the usage of the $\Psi$-operator
instead of the $\Lambda$-operator. 

If we insert Eq.~\ref{Rop} into Eq.~\ref{REQ} we obtain the following system for the new 
population densities:
\begin{multline}
    \sum_{j<i} n_{j,\text{new}} [R_{ji}^{*}][n_{\text{new}}]\\
   -n_{i,\text{new}}\left\{\sum_{j<i} \left(\div{n_j^{*}}{n_i^{*}}\right)
		[R_{ij}^{*}][n_{\text{new}}] 
	     +\sum_{j>i} 
		 [R_{ij}^{*}][n_{\text{new}}]\right\}  \\
   +\sum_{j>i} n_{j,\text{new}} 
     \left(\div{n_i^{*}}{n_j^{*}}\right)[R_{ji}^{*}][n_{\text{new}}]  \\
    +\sum_{j<i} n_{j,\text{new}} \left(\DRji[n_{\text{old}}]+C_{ji}\right)   \\
   -n_{i,\text{new}}\left\{\sum_{j<i} \left(\div{n_j^{*}}{n_i^{*}}\right)
		\left(\DRij[n_{\text{old}}]+C_{ji}\right) \right.               \\ 
   \left. +\sum_{j>i} 
		 \left(\DRij[n_{\text{old}}]+C_{ij}\right)\right\}      \\
   +\sum_{j>i} n_{j,\text{new}} 
     \left(\div{n_i^{*}}{n_j^{*}}\right)\left(\DRji[n_{\text{old}}]+C_{ij}\right)
     = 0. \label{OSREQ}
\end{multline}
Due to its 
construction, the \Rijstar-operator contains 
information about the influence of a particular level
on {\em all} radiative transitions. Therefore, we are able to treat the 
complete multi-level
NLTE radiative transfer problem including active continua and overlapping lines. 
The $[E(\lambda)]$-operator, at the same time, gives us information about the
strength of the coupling of a radiative transition to all levels considered. This
information may be used to include or neglect certain couplings {\em dynamically}
during the iterative solution of Eq.~\ref{OSREQ}.
For example, one could include all possible couplings in the first iteration,
and use the relative magnitudes of the  $[E(\lambda)]$'s to decide which couplings to
include in subsequent iterations (and repeat this process after each set of iterations).
 Furthermore, we have not
yet specified either a method for the formal solution of the radiative transfer
equation or a method for the construction of the approximate $\L$-operator
(and, correspondingly, the $\Rijstar$-operator).
Here, we use the $\Lstar$ operator constructed in Paper I. Due to storage
considerations, we can  only use the diagonal part of the 3D $\Lstar$ in
the calculations discussed below.
However, any method for the formal
solution of the radiative transfer equation and the construction of the ALO may
be used.

\subsection{Iterative solution}

The system Eq.~\ref{OSREQ} for $[n_{\text{new}}]$ is non-linear with respect to 
the $n_{i,\text{new}}$ and $n_e$ because the coefficients of the $\Rijstar$ and
$\Rjistar$-operators are quadratic in $n_{i,\text{new}}$ and of the
dependence the 
Saha-Boltzmann factors and the collisional rates on the electron density,
respectively. 
The system is closed by the abundance and charge conservation equations. 
To simplify the iteration scheme, and to take advantage of the fact that 
not all levels strongly influence all radiative transitions, we use a 
linearized and splitted iteration scheme for the solution of Eq.~\ref{OSREQ}. 
This scheme has the further advantage that
many different elements in different ionization stages and even molecules
can be treated consistently.  Problems where this is important are, e.g., the
modeling of nova and supernova photospheres or cool stellar
atmospheres, where one typically  
finds very large temperature gradients within the line forming region of the atmosphere. 

  First, we follow \citep{rh91} and replace terms of the form
$n_{j,\text{new}} [R_{ji}^{*}][n_{\text{new}}]$ in Eq.~\ref{OSREQ} by
$n_{j,\text{old}} [R_{ji}^{*}][n_{\text{new}}]$: 
\begin{multline}
    \sum_{j<i} n_{j,\text{old}} [R_{ji}^{*}][n_{\text{new}}] \\
   - n_{i,\text{old}}\left\{\sum_{j<i}\left(\div{n_j^{*}}{n_i^{*}}\right)
		[R_{ij}^{*}][n_{\text{new}}]\
	     +\sum_{j>i} 
		 [R_{ij}^{*}][n_{\text{new}}]\right\}     \\
   +\sum_{j>i} n_{j,\text{old}} 
     \left(\div{n_i^{*}}{n_j^{*}}\right)[R_{ji}^{*}][n_{\text{new}}] 
    +\sum_{j<i} n_{j,\text{new}} \left(\DRji[n_{\text{old}}]+C_{ji}\right) \\
   -n_{i,\text{new}}\left\{\sum_{j<i}\left(\div{n_j^{*}}{n_i^{*}}\right)
		\left(\DRij[ n_{\text{old}}]+C_{ij}\right)  \right. \\
	     \left. +\sum_{j>i} 
		 \left(\DRij[n_{\text{old}}]+C_{ij}\right)\right\}      \\
   +\sum_{j>i} n_{j,\text{new}}
     \left(\div{n_i^{*}}{n_j^{*}}\right)\left(\DRji[n_{\text{old}}]+C_{ji}\right)
     = 0. \label{itREQ}
$$
\end{multline}

This removes the major part of the non-linearity of Eq.~\ref{OSREQ} but the modified
system is still non-linear with respect to $n_e$ and still has the high
dimensionality of the original system. However, not all levels are strongly
coupled to all other levels and not all elements depend strongly on the rates
of other elements.  Therefore, we may make the additional assumption that $N_{\kappa,\rm
old}\approx N_{\kappa,\text{new}}$, where the index $\kappa$ refers to
the ground state of the next higher ionization stage, and all collisional rates are evaluated using
the current value of $n_e$ for the solution of the rate
equations at a given iteration.  These approximations close the rate-equations,
either ion by ion or  element by element, and a separate solution of the charge
conservation constraint is possible.  This will first slow the iteration process,
especially if the electron density changes considerably during the initial
iterations, but in the convergence limit it will be accurate.

With this iteration scheme, Eq.~\ref{itREQ} can be solved for each ion or element {\em separately} 
if the electron density is given. However, if transitions between two ions or elements are strongly
coupled, we can easily combine the sets of equations and solve them simultaneously in order to
include these couplings directly in the iterations. The most important 
 advantage of Eq.~\ref{itREQ} is that
it is {\em linear} for a given $n_e$ and thus, in general, its
solution is more stable and  
uses much less computer resources (time and memory)
than the direct solution of the original non-linear equations.

 We have assumed so far that the electron density $n_e$ is given. However,
although this is a good assumption if only trace elements are considered in
NLTE, the electron density may, in certain regions of the temperature vs.\ gas
pressure plane, be very sensitive to NLTE effects. This can be taken into
account by using either a fixed point iteration scheme for the electron density
or, in particular, if many species or molecules are included in the NLTE
equation of state, by a modification of the LTE partition functions to include
the effects of NLTE in the ionization equilibrium. The latter method
replaces the partition function, $Q=\sum g_i\exp(-\chi_i/kT)$, with its NLTE
generalization, $Q_{\text{NLTE}} = \sum b_i g_i\exp(-\chi_i/kT)$, and uses $Q_{\rm
NLTE}$ in the solution of the ionization/dissociation equilibrium equation. In
this paper, we use this method because of the potentially large number of
elements and ionization stages included in the ionization equilibrium (and not
all of them in NLTE).  We could solve Eq.~\ref{itREQ} directly, bypassing the
additional splitting of the iteration, as a system of non-linear equations for
the electron density and the updated population densities. This may be
favorable under certain conditions, e.g., if the electron density is strongly
influenced by NLTE effects.   However,  solving large non-linear sets
of equations is time consuming, complex and error prone.  In practical tests we
found that the method described above to be very reliable, which is an
important advantage in the long running 3D calculation.

Our iteration scheme for the solution of the multi-level NLTE problem can be
summarized as follows: (1) for given $n_i$ and $n_e$, solve the radiative
transfer equation at each wavelength point and update the radiative rates and
the approximate rate operator, (2) solve the linear system, Eq.~\ref{itREQ}, for
each group of ions or elements for a given electron density, (3)  after all
rate equations have been solved, compute new
electron densities (by either fixed point iteration or the generalized
partition function method  using the new departure coefficients estimates to update
the ionization equilibria).  Updating the $n_e$ and iterating only steps (2)+(3) 
will lead to convergence problems as the data going into the rate equations are
sensitive to $n_e$.  The iterations are repeated until a prescribed
accuracy for the $n_e$ and the $n_i$ is reached.  This method 
gives a fully consistent converged solution for the $n_i$ and $n_e$.

\subsection{3D details}

A major issue for 3D NLTE calculations are the memory requirements for storing
the relevant data. For each voxel we need to store (at least) the $n_i$, line
profiles, the radiative rates (up/down), the rate operators (one up/down pair
per considered interaction), general equation of state data (partial pressures
for all species), and the data needed for the solution of the 3D radiative
transfer equation at every wavelength point (re-usable, only the current
wavelength point needs to be stored). For the smallest test case discussed
below with 70,785 voxels, 62 NLTE levels (H I, He I+II), 576 transitions, 913
explicit coupled transitions for the $\Rijstar$ operators and a total of 894
species in the equation of state this results in a relatively small footprint
of 2.8GB total. However, for a large case with 274,625 voxels, 4686 NLTE
levels, 81,652 transitions, and 165,063 explicit coupled transitions for the
$\Rijstar$ operators (this roughly corresponds to typical \phxO\ NLTE models)
the total memory footprint increases to about 1.2TB. Whereas the small case
could be handled on a single CPU core, larger cases require a domain
decomposition method and distributed memory on large scale parallel computers.
Therefore, we have implemented the NLTE/3D module with a hierarchical domain
decomposition (DD) method that distributes the 
equation of state (EOS) data, NLTE level, radiative rates and rate operator
data over a group of processes  (a 'DD group') so that each
process  of a DD group only holds  this data for a
fraction of the voxels.   For example, if the size of DD group is
$n$ MPI processes, then each process of that DD group stores the EOS data and
the NLTE data (populations, rates, rate operators) for 
for a fraction $1/n$ of all the voxels in the DD group. Each
process in a  DD group holds all the data  required
for the solution of the 3DRT problem  at any given wavelength,
the 3DRT solution is parallelized  over solid angles within a
 DD group (this works as the storage requirements for 3DRT are
small compared to the NLTE and the EOS requirements),  cf.\
\cite{3drt_paper1}.  We then use sets of DD groups to handle
different sets of wavelength points to build up partial radiative rates and
rate operators.  For example, if we have $m$ DD groups (for a total of $n\times
m$ processes) each DD group will work on $1/m$ of the overall wavelength
points The results  from each DD group are then combined to build
up the rates and operators before Eq.~\ref{itREQ} is solved, also distributed
over  MPI processes. This scheme can also be used to optimize
communication between processes, e.g., by mapping domain decomposition groups
on single compute nodes with shared memory (for small cases).  The
communication between processes in different  DD groups is
localized  by grouping together the partial rates and operators
(in the simplest case a simple {\tt MPI\_allreduce} between processes with the
same domain but different  DD groups) and to distribute the
results of the solution of Eq.~\ref{itREQ}  for any voxel to the
different DD groups that need the data for this voxel.  Depending
on the number of voxels and the number of NLTE levels and transitions as well
as the capacity of the parallel computer used, the size $n$ of a DD group was
for the  models shown here between 96 and 480 (the theoretical maximum is
limited by the number of voxels, here about 66000) and the number of $m$ such
DD groups was between 140 and 512 (here the theoretical limit is set by the
number of wavelength points, which is around 500000 for the largest model shown
below). This means that the 3D NLTE calculations can scale up to a very large
number of processes, for the calculations reported below we have used up to
67200 processes, production simulations could easily use several million
processes.

The CPU time requirements are also significant. On current Intel Xeon E5420
CPUs with 2.50GHz clock-speed, the time for a single iteration (solution of the
3D radiative transfer for all wavelength points, solution of Eq.~\ref{itREQ}
for all species, etc) on a single core for the small test case would be about
0.5 years (or 4400 hours or $3.6\alog{-11}\,$Hubble where $1\,$Hubble is
the work that a single CPU core could do in 13.8\,Gyr). For the large case we
estimate a single-core CPU time of about 4300 years ($3\alog{-7}\,$Hubble) for
a single iteration.  Clearly, running even the smallest test case on a serial
computer is impractical.  However, on a parallel computer with 4096 (MPI)
processes, a single iteration for the small test case requires only about 3800
seconds (1.05 hours, actual time measured on the HLRN-II SGI ICE-2 system), with
about 45 iterations required for convergence, this corresponds to about 50
hours wallclock time (about 200kh single core CPU time). The large case would
require about 215kyr ($1.5\alog{-5}\,$Hubble or $15\,\mu$Hubble) serial
CPU time, however, as this case could scale easily up 8 million cores or more, the
wallclock time could be as low as about 1 month for the full calculation
(again, without using any simplification or approximation).

\section{Results}

In order to verify the NLTE/3D module, we use simple test models. We use the
temperature-pressure structure of a \phxO\  spherically symmetric
stellar atmosphere model with $\Teff = 9800\,$K, $\logg=4.5$ and solar
abundances.  This was mapped to the spherical coordinate system
mode of \phxT, this is the same procedure that we used in \cite{3drt_paper6}.
To simplify the calculations, we perform the test calculations without any
background LTE lines (this saves significant computer time). With this test
structure, we can compute fully comparable 1D and 3D models to verify if the 3D
NLTE module is working correctly.

In the first test, we solve the multi-level 3D NLTE problem in a spherical
coordinate system  with $65$ radial points and $33$ points each in $\theta$ and
$\phi$, for a total of $66625$ non-vacuum voxels and for H I (30 level), He I
(19 levels) and He II (10 levels) model atoms. This model has a total of 61
levels, 517 lines, 1002 rate operators (lines and continua) and uses 18708
wavelength points to model lines and continua.  All lines are considered with
depth dependent Voigt profiles (Stark profiles for H I).
 
\begin{figure*}
\centering
\begin{minipage}{0.45\hsize}
\resizebox{\hsize}{!}{\includegraphics[angle=00]{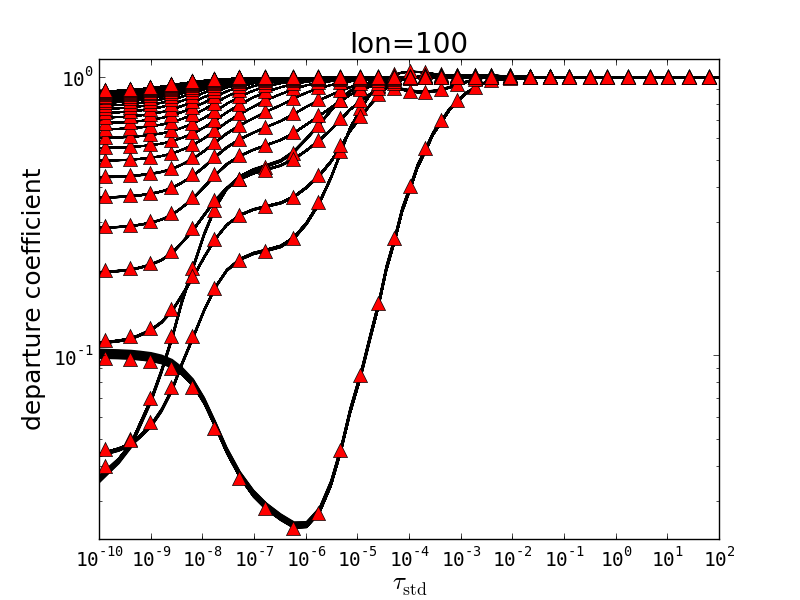}}
\end{minipage}
\begin{minipage}{0.45\hsize}
\resizebox{\hsize}{!}{\includegraphics[angle=00]{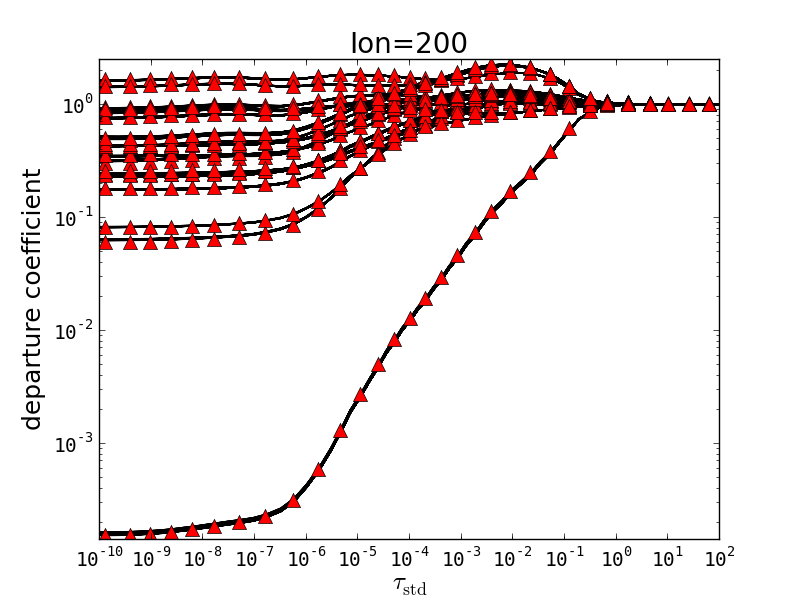}}
\end{minipage}
\begin{minipage}{0.45\hsize}
\resizebox{\hsize}{!}{\includegraphics[angle=00]{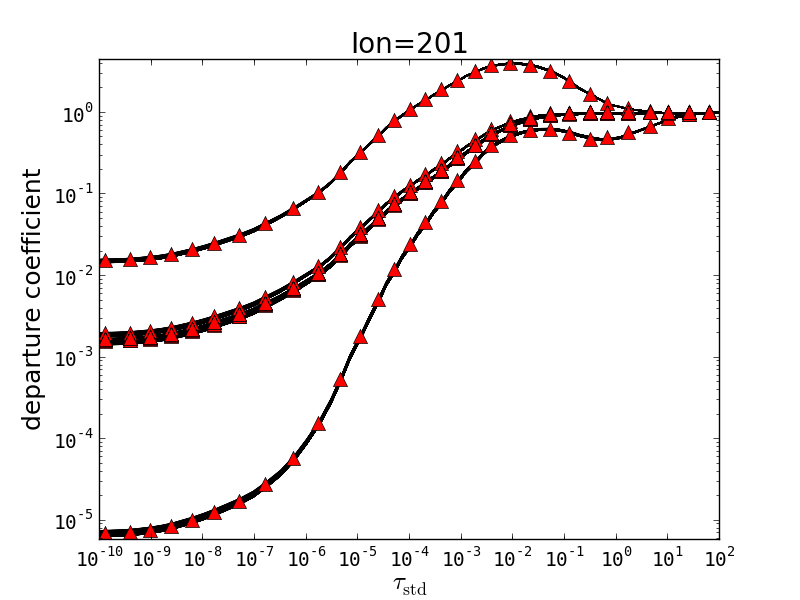}}
\end{minipage}
\caption{\label{fig:bi}
For our first test with NLTE treatment of H~I and He~I--II, the departure coefficients $b_i$,
are shown.
The red symbols show the results of the 1D calculation
whereas the black lines show the results of the 3D calculation for all voxels.
}
\end{figure*}

In Fig.~\ref{fig:bi} we display the departure coefficients $b_i \equiv
n_i/n_i^*$ for H~I,
He~I and He~II. The (red) symbols show the results of the 1D calculation
whereas the black lines show the results of the 3D calculation for all voxels.
The spread of the black lines is due to the limited solid angle resolution used
in the 3D test run \citep[see][for details]{3drt_paper6}.  The agreement is
excellent, only for the outermost layers with $\tau_{1.2\mu} < 10^{-5}$ there
are small differences between the 1D and 3D departure coefficients for the
lower levels. These differences could be caused by the small $(\theta,\phi)$
voxel resolution of the 3D test \citep[the $r$ resolution is nearly
identical to the 
1D model, see][for a similar effect]{3drt_paper10}. 

\begin{figure*}
\centering
\resizebox{\hsize}{!}{\includegraphics[angle=00]{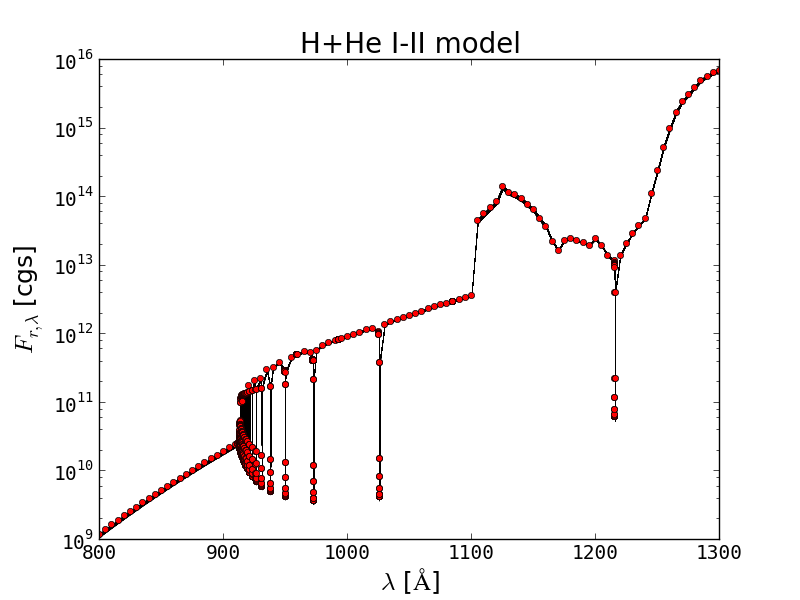}}
\caption{\label{fig:Lyman}
For our first test with NLTE treatment of H~I and He~I--II the
spectral region near the Lyman break is shown. 
The red lines show the results of the 1D NLTE calculation
whereas the black lines show the maximum and minimum of the radial component 
of the flux vector over all outermost voxels from the 3D calculation.
The spread in the 3D model is due to the finite numerical resolution,
see \cite[][paper VI]{3drt_paper6}.
}
\end{figure*}

\begin{figure*}
\centering
\resizebox{\hsize}{!}{\includegraphics[angle=00]{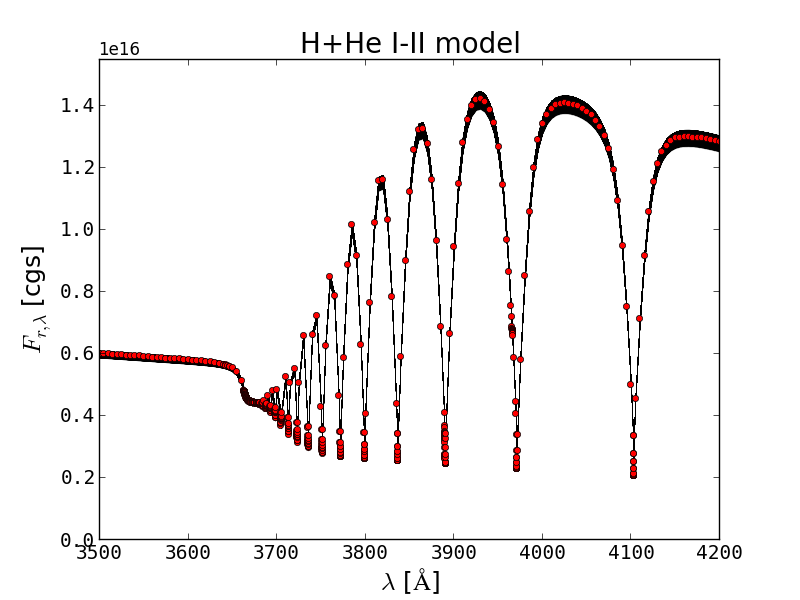}}
\caption{\label{fig:Balmer}
For our first test with NLTE treatment of H~I and He~I--II the
spectral region near the Balmer jump is shown. 
The red lines show the results of the 1D NLTE calculation
whereas the black lines show the maximum and minimum of the radial component 
of the flux vector over all outermost voxels from the 3D calculation.
The spread in the 3D model is due to the finite numerical resolution,
see paper VI.
}
\end{figure*}

\begin{figure*}
\centering
\resizebox{\hsize}{!}{\includegraphics[angle=00]{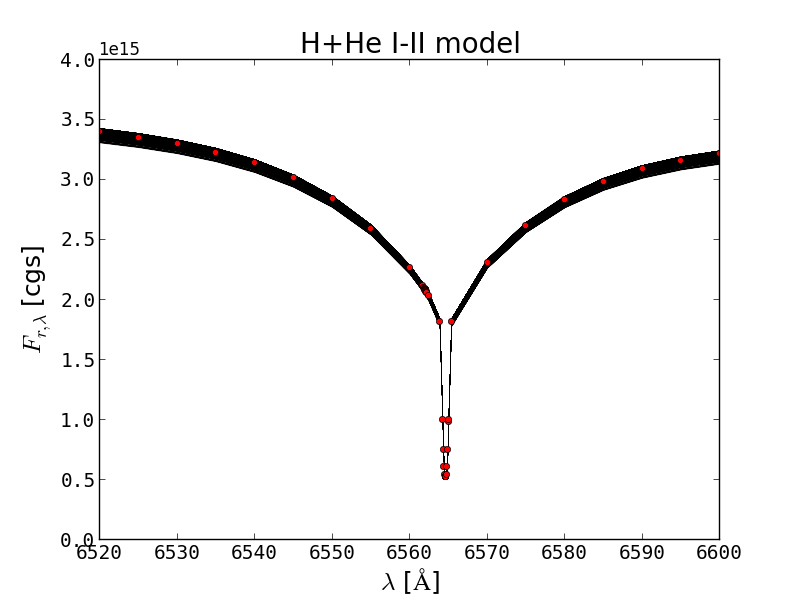}}
\caption{\label{fig:Halpha}
For our first test with NLTE treatment of H~I and He~I--II the
spectral region near $H_\alpha$ is shown. 
The red lines show the results of the 1D NLTE calculation
whereas the black lines show the maximum and minimum of the radial component 
of the flux vector over all outermost voxels from the 3D calculation.
The spread in the 3D model is due to the finite numerical resolution,
see paper VI.
}
\end{figure*}

The spectra in the regions around the Lyman and Balmer jumps and $H_\alpha$ are
compared in 
Figs.~\ref{fig:Lyman} -- \ref{fig:Halpha},
where the red symbols are again the 1D fluxes and the black lines are the
radial components of the flux vectors at the surface voxels of the 3D model
(the wavelength resolution was  taken directly from the NLTE iterations).  The
spectra are nearly identical, the ``bandwidth'' in the 3D spectra is again the
resolution effect discussed in \citet{3drt_paper6}. 
 In the wings of the Balmer lines, the resolution spread in 
the flux results in differences of about 5\% to the the 1D model. To reduce 
the spread by 1/2, we estimate based on the results of Paper VI that about 
4 times more solid angle points are needed, which is no problem for
full production calculations.
Note that only the NLTE lines and continua and LTE background
continua are included in the modeling (to save time), no LTE background lines
are considered. 

The second test uses the same basic setup as discussed above, but includes a
full NLTE treatment for H and He and the first 3 ionization stages of C, N, O,
and Mg. In detail, the setup is identical for H and He and we use C~I (230
levels), C~II (85 levels), C~III (79 levels), N~I(254 levels), N~II (152 levels),
N~III(87 levels), O~I (146 levels), O~II (171 levels), O~III (137 levels), Mg~I
(179 levels), Mg~II (74 levels), and Mg~III (90 levels), for a total of 21 ions,
1749 levels, and 15478 line transitions.  As before we use an identical setup
for the 1D comparison calculations.

\begin{figure*}
\centering
\begin{minipage}{0.45\hsize}
 \resizebox{\hsize}{!}{\includegraphics[angle=00]{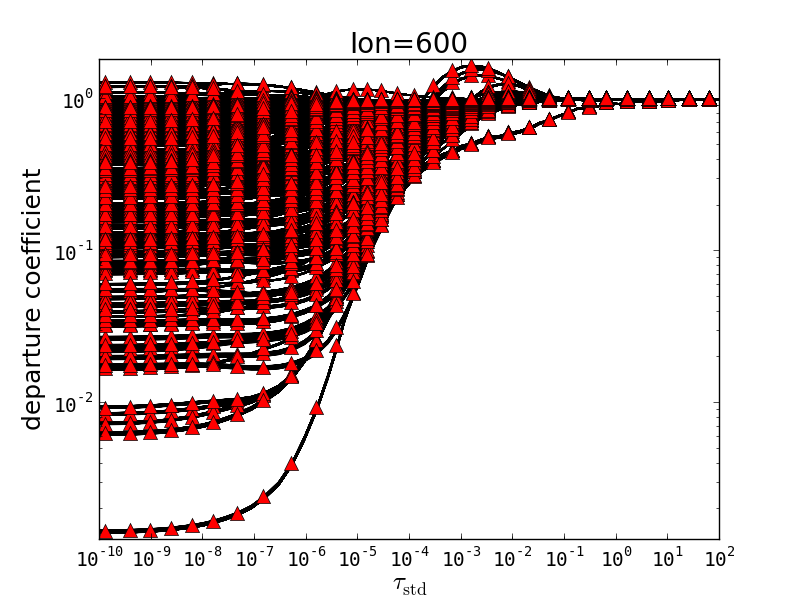}}
\end{minipage}
\begin{minipage}{0.45\hsize}
 \resizebox{\hsize}{!}{\includegraphics[angle=00]{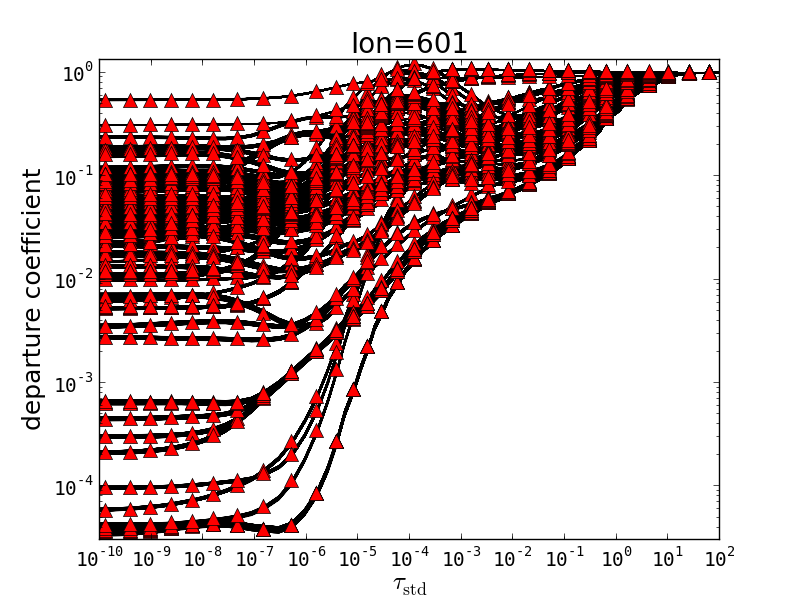}}
\end{minipage}
\begin{minipage}{0.45\hsize}
 \resizebox{\hsize}{!}{\includegraphics[angle=00]{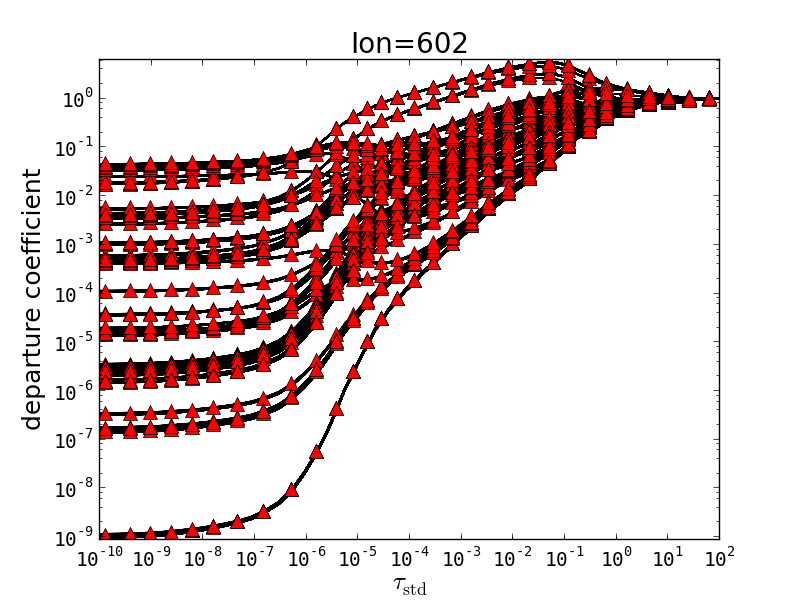}}
\end{minipage}
\caption{\label{fig:HHeCNOMg:bi_600}
For our second test case with NLTE treatment of H~I, He~I--II,
C~I--III, N~I---III, O~I---III, and Mg~I---III the
departure coefficients, $b_i$, for C~I---III are shown.
The red symbols show the results of the 1D calculation
whereas the black lines show the results of the 3D calculation for all voxels.
}
\end{figure*}

\begin{figure*}
\centering
\begin{minipage}{0.45\hsize}
 \resizebox{\hsize}{!}{\includegraphics[angle=00]{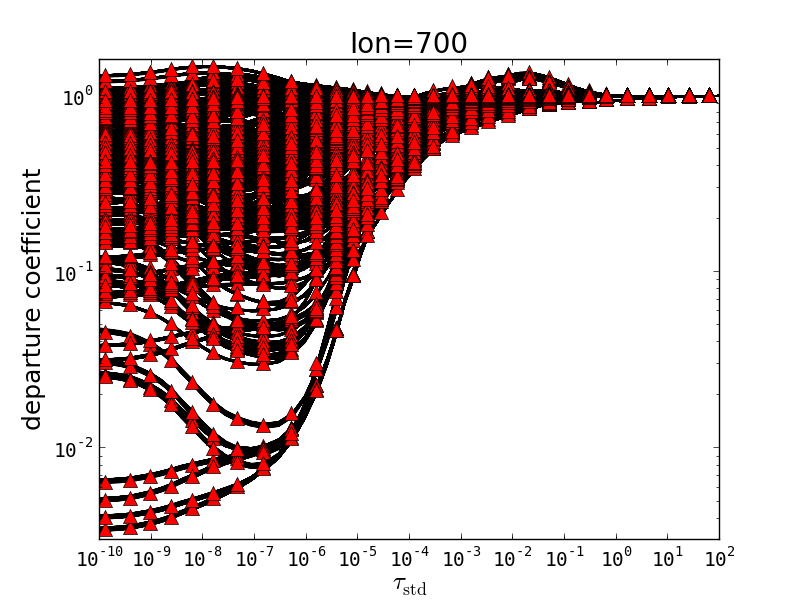}}
\end{minipage}
\begin{minipage}{0.45\hsize}
 \resizebox{\hsize}{!}{\includegraphics[angle=00]{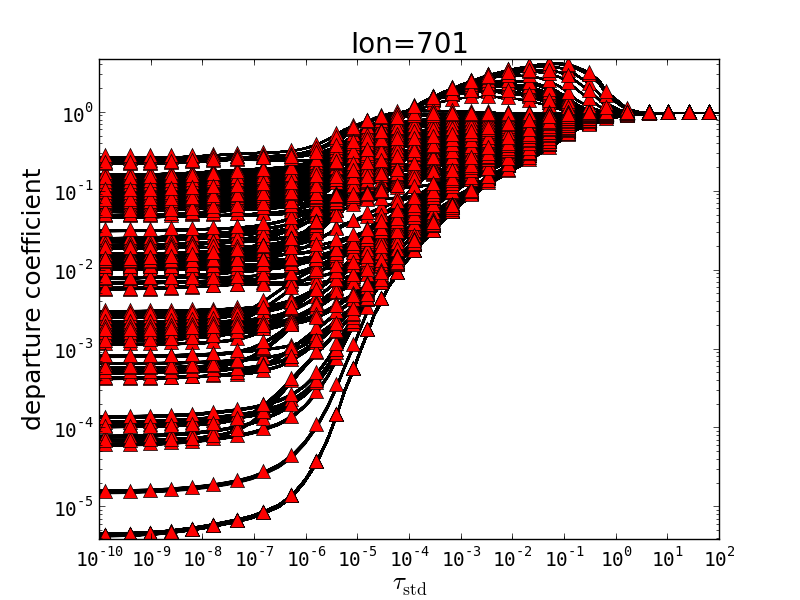}}
\end{minipage}
\begin{minipage}{0.45\hsize}
 \resizebox{\hsize}{!}{\includegraphics[angle=00]{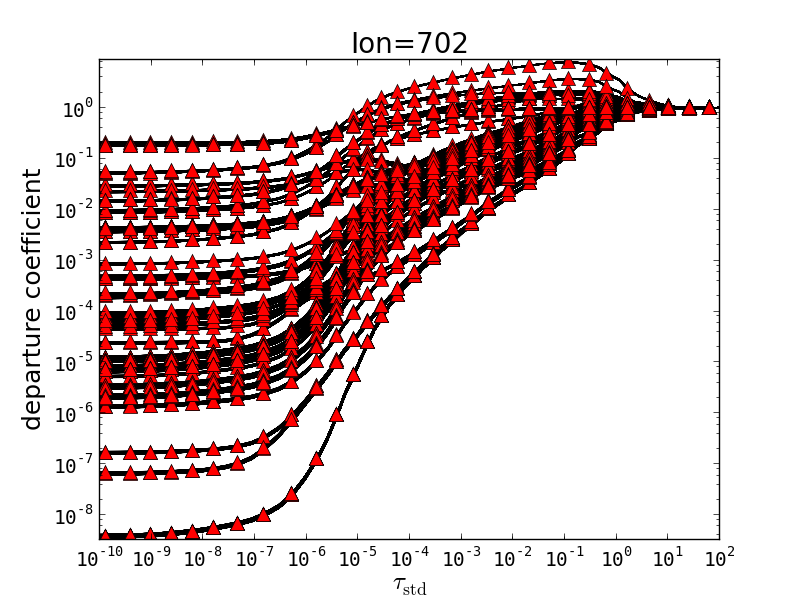}}
\end{minipage}
\caption{\label{fig:HHeCNOMg:bi_700}
For our second test case with NLTE treatment of H~I, He~I--II,
C~I--III, N~I---III, O~I---III, and Mg~I---III the
departure coefficients, $b_i$, for N~I---III are shown.
The red symbols show the results of the 1D calculation
whereas the black lines show the results of the 3D calculation for all voxels.
}
\end{figure*}

\begin{figure*}
\centering
\begin{minipage}{0.45\hsize}
 \resizebox{\hsize}{!}{\includegraphics[angle=00]{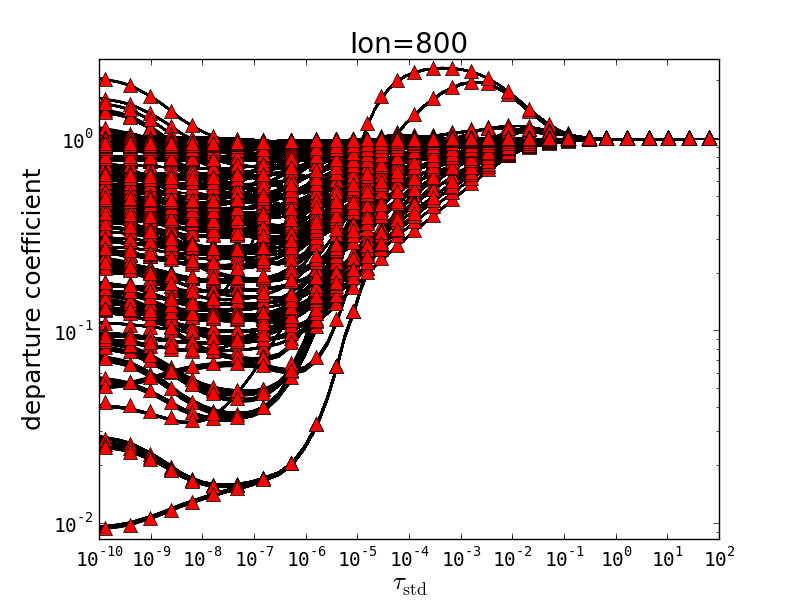}}
\end{minipage}
\begin{minipage}{0.45\hsize}
 \resizebox{\hsize}{!}{\includegraphics[angle=00]{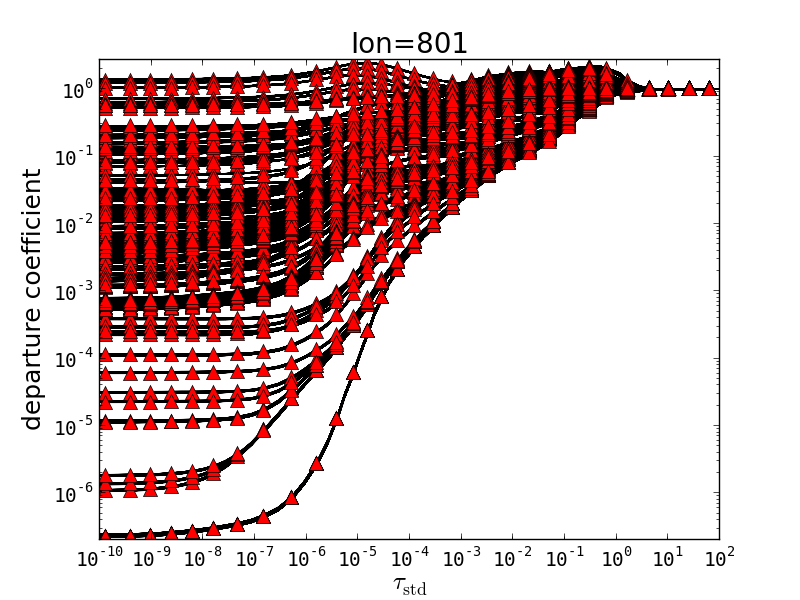}}
\end{minipage}
\begin{minipage}{0.45\hsize}
 \resizebox{\hsize}{!}{\includegraphics[angle=00]{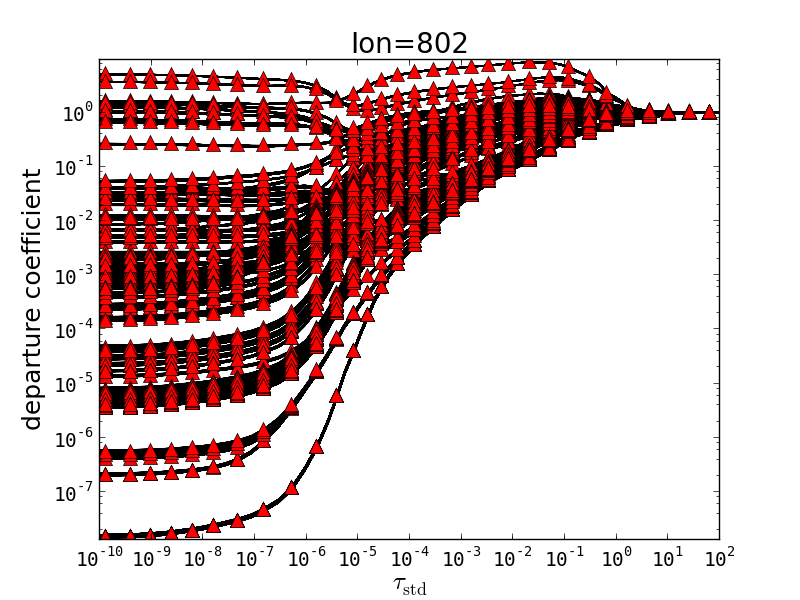}}
\end{minipage}
\caption{\label{fig:HHeCNOMg:bi_800}
For our second test case with NLTE treatment of H~I, He~I--II,
C~I--III, N~I---III, O~I---III, and Mg~I---III the
departure coefficients, $b_i$, for O~I---III are shown.
The red symbols show the results of the 1D calculation
whereas the black lines show the results of the 3D calculation for all voxels.
}
\end{figure*}

\begin{figure*}
\centering
\begin{minipage}{0.45\hsize}
 \resizebox{\hsize}{!}{\includegraphics[angle=00]{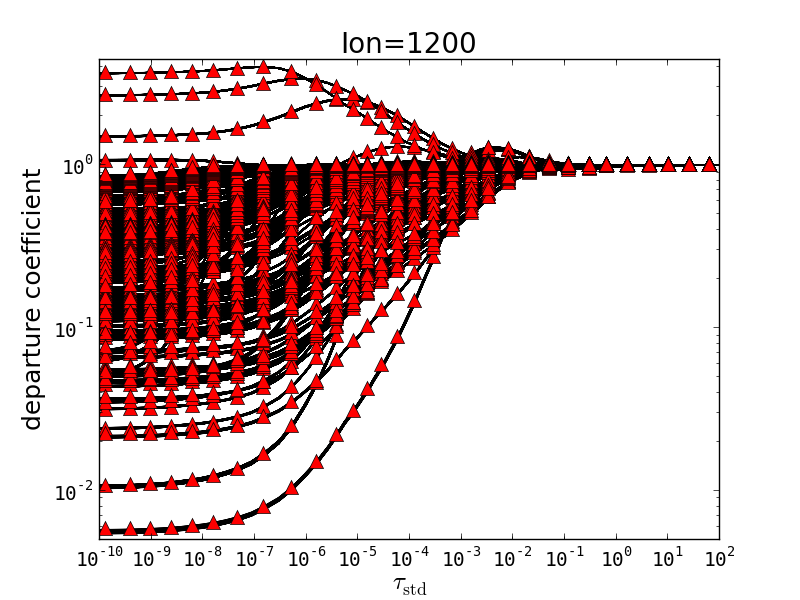}}
\end{minipage}
\begin{minipage}{0.45\hsize}
 \resizebox{\hsize}{!}{\includegraphics[angle=00]{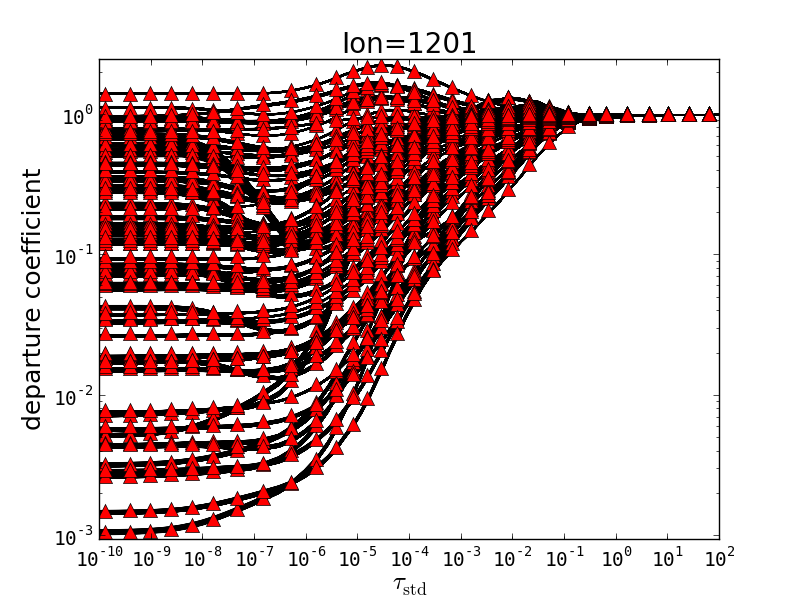}}
\end{minipage}
\begin{minipage}{0.45\hsize}
 \resizebox{\hsize}{!}{\includegraphics[angle=00]{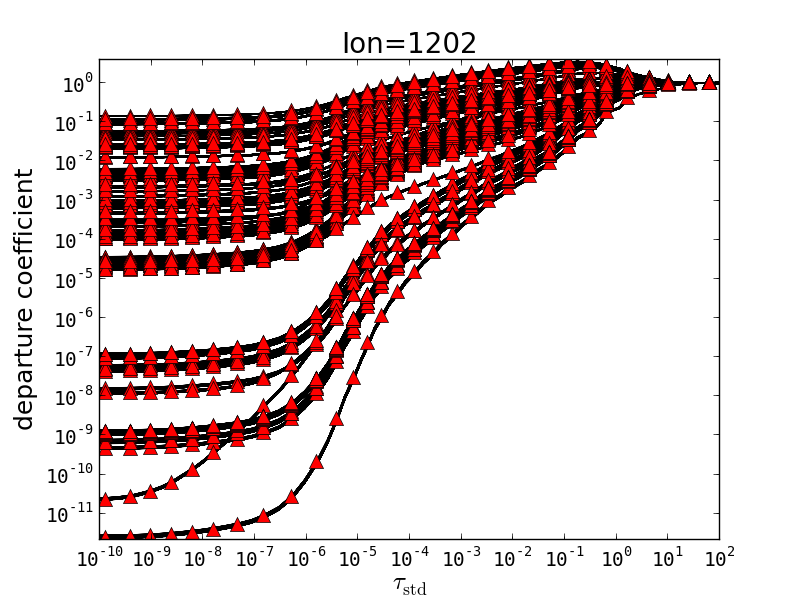}}
\end{minipage}
\caption{\label{fig:HHeCNOMg:bi_1200}
For our second test case with NLTE treatment of H~I, He~I--II,
C~I--III, N~I---III, O~I---III, and Mg~I---III the
departure coefficients, $b_i$, for Mg~I---III are shown.
The red symbols show the results of the 1D calculation
whereas the black lines show the results of the 3D calculation for all voxels.
}
\end{figure*}

\begin{figure*}
\centering
\resizebox{\hsize}{!}{\includegraphics[angle=00]{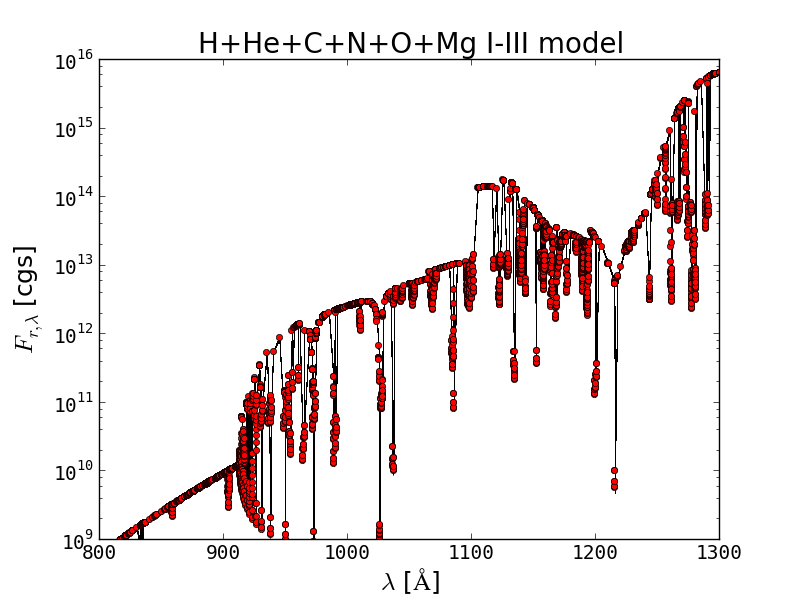}}
\caption{\label{fig:HHeCNOMg:Lyman_HHeCNOMg}
For our second test case with NLTE treatment of H~I, He~I--II,
C~I--III, N~I---III, O~I---III, and Mg~I---III the
spectral region near the Lyman break is shown.
The red lines show the results of the 1D NLTE calculation
whereas the black lines show the maximum and minimum of the radial component 
of the flux vector over all outermost voxels from the 3D calculation.
The spread in the 3D model is due to the finite numerical resolution,
see paper VI.
}
\end{figure*}

\begin{figure*}
\centering
\resizebox{\hsize}{!}{\includegraphics[angle=00]{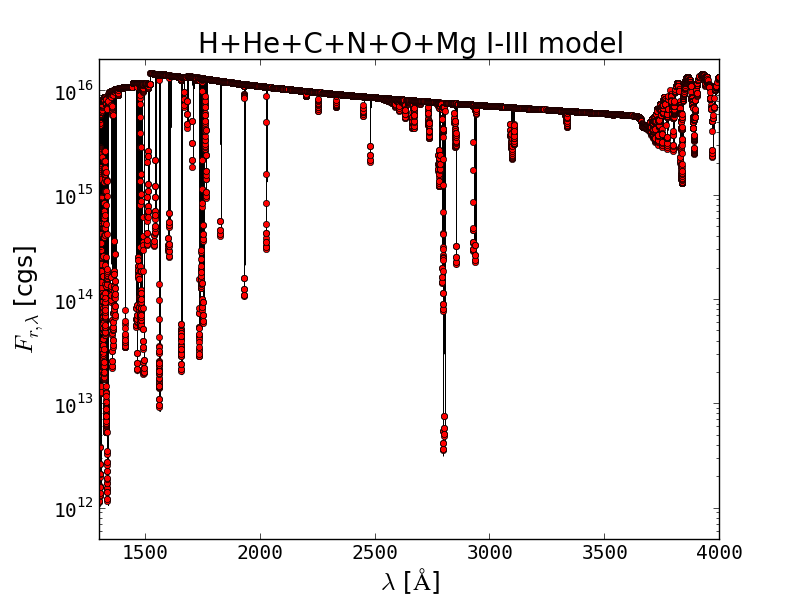}}
\caption{\label{fig:HHeCNOMg:IUE_HHeCNOMg}
For our second test case with NLTE treatment of H~I, He~I--II,
C~I--III, N~I---III, O~I---III, and Mg~I---III the spectral region in the near UV is shown.
The red lines show the results of the 1D NLTE calculation
whereas the black lines show the maximum and minimum of the radial component 
of the flux vector over all outermost voxels from the 3D calculation.
The spread in the 3D model is due to the finite numerical resolution,
see paper VI.
}
\end{figure*}

Some of the results for the departure coefficients are shown in examples in
Figs. \ref{fig:HHeCNOMg:bi_600} -- \ref{fig:HHeCNOMg:bi_1200}.  The comparison
to the equivalent 1D model is of the same quality as in the simpler test case
shown above. The plots show only every other point from the 1D model
to reduce clutter. Even complex behavior in the departure coefficients is well
reproduced in the 3D model. The deviations from the 1D model are smaller than
the width of the bands produced by the numerical resolution of the solid angle
grid \citep[see][]{3drt_paper6}. Similar results hold for the spectra, shown for
two examples in Figs.~\ref{fig:HHeCNOMg:Lyman_HHeCNOMg} and
\ref{fig:HHeCNOMg:IUE_HHeCNOMg}.

This test model has about 111,000 wavelength points (about 10
times more than the small test case) and requires about 706GB to store the
full spectral data (mean intensities and flux vectors for all voxels and
wavelengths for detailed analysis), storing just the  outer spectrum (flux
vectors) for plotting requires about 5GB, the departure coefficients and
occupation number densities require about 2.7GB storage for all voxels. The
larger test was run on a Cray XE30 supercomputer using 49248 MPI processes
(2052 nodes) using about 970MB RAM per process and 2680 seconds wallclock per
iteration.  For comparison, the 1D equivalent model uses about 43s per
iteration with 48 MPI processes on the same machine.

\begin{figure*}
\centering
\begin{minipage}{0.45\hsize}
 \resizebox{\hsize}{!}{\includegraphics[angle=00]{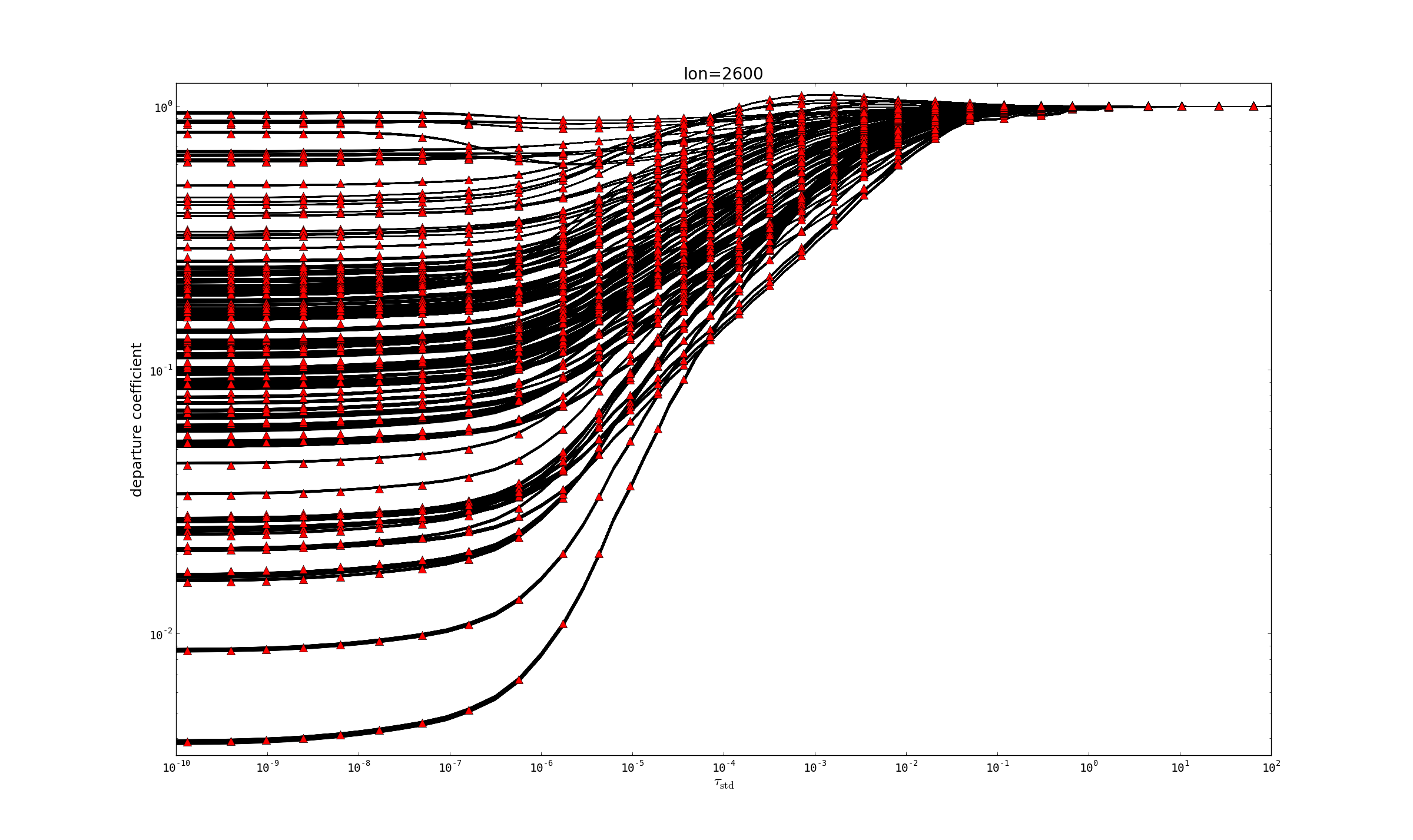}}
\end{minipage}
\begin{minipage}{0.45\hsize}
 \resizebox{\hsize}{!}{\includegraphics[angle=00]{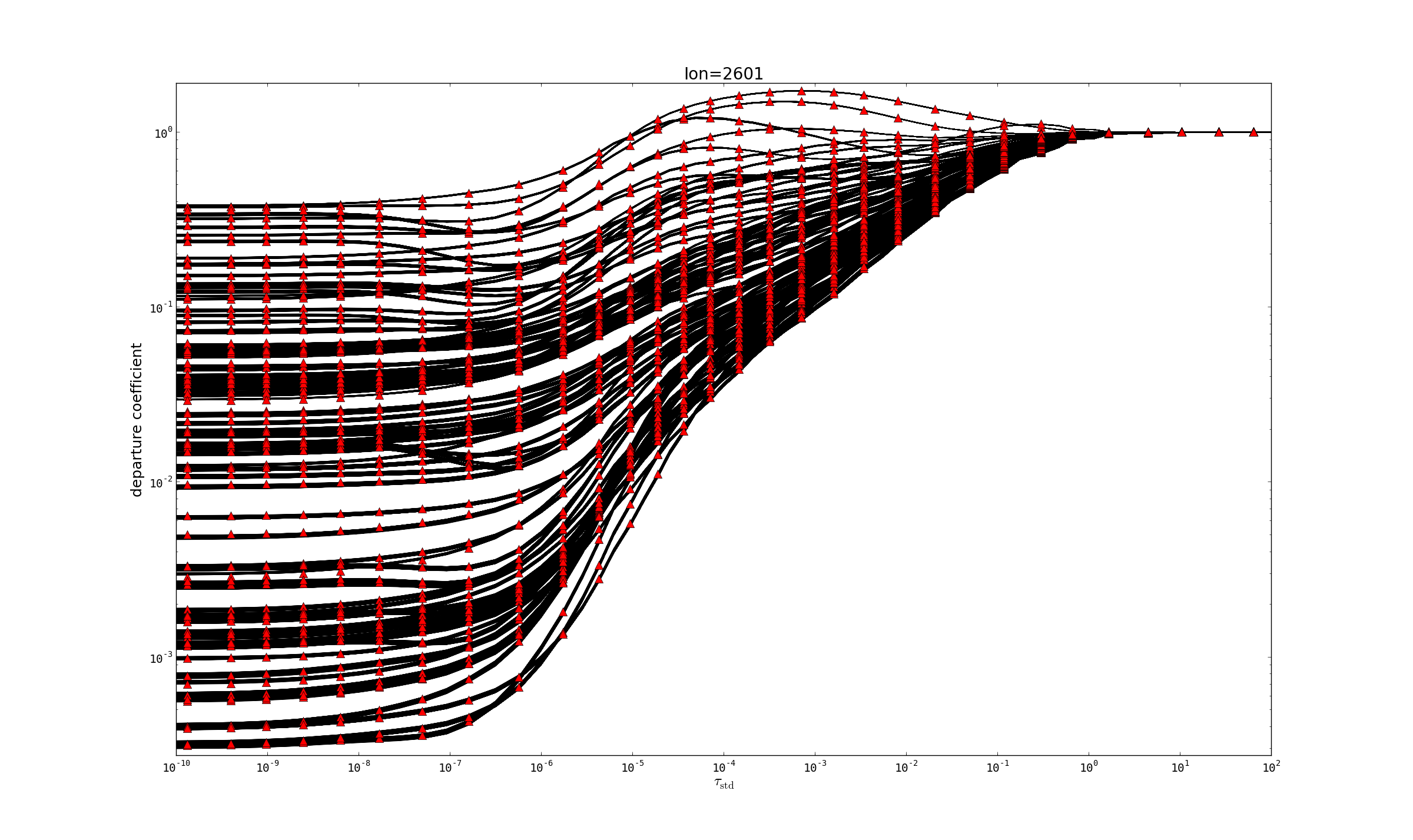}}
\end{minipage}
\begin{minipage}{0.45\hsize}
 \resizebox{\hsize}{!}{\includegraphics[angle=00]{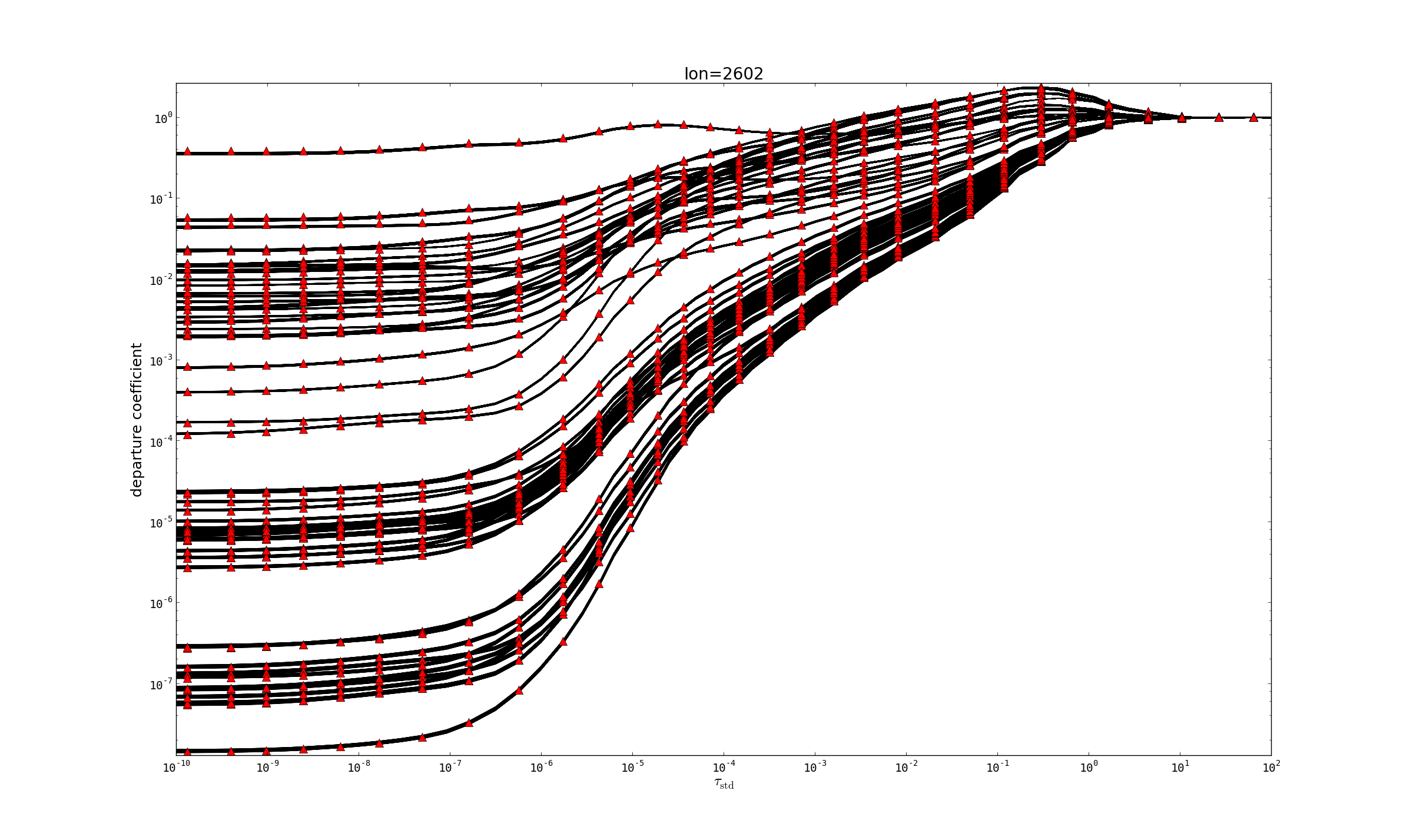}}
\end{minipage}
\caption{\label{fig:HHeFe:bi_2600}
For our third test case with NLTE treatment of H~I, He~I--II,
and Fe~I--III,  the
departure coefficients, $b_i$, for Fe~I---III are shown. For clarity,
we plot only every 10th level.
The red symbols show the results of the 1D calculation
whereas the black lines show the results of the 3D calculation for all voxels.
}
\end{figure*}

\begin{figure*}
\centering
\resizebox{\hsize}{!}{\includegraphics[angle=00]{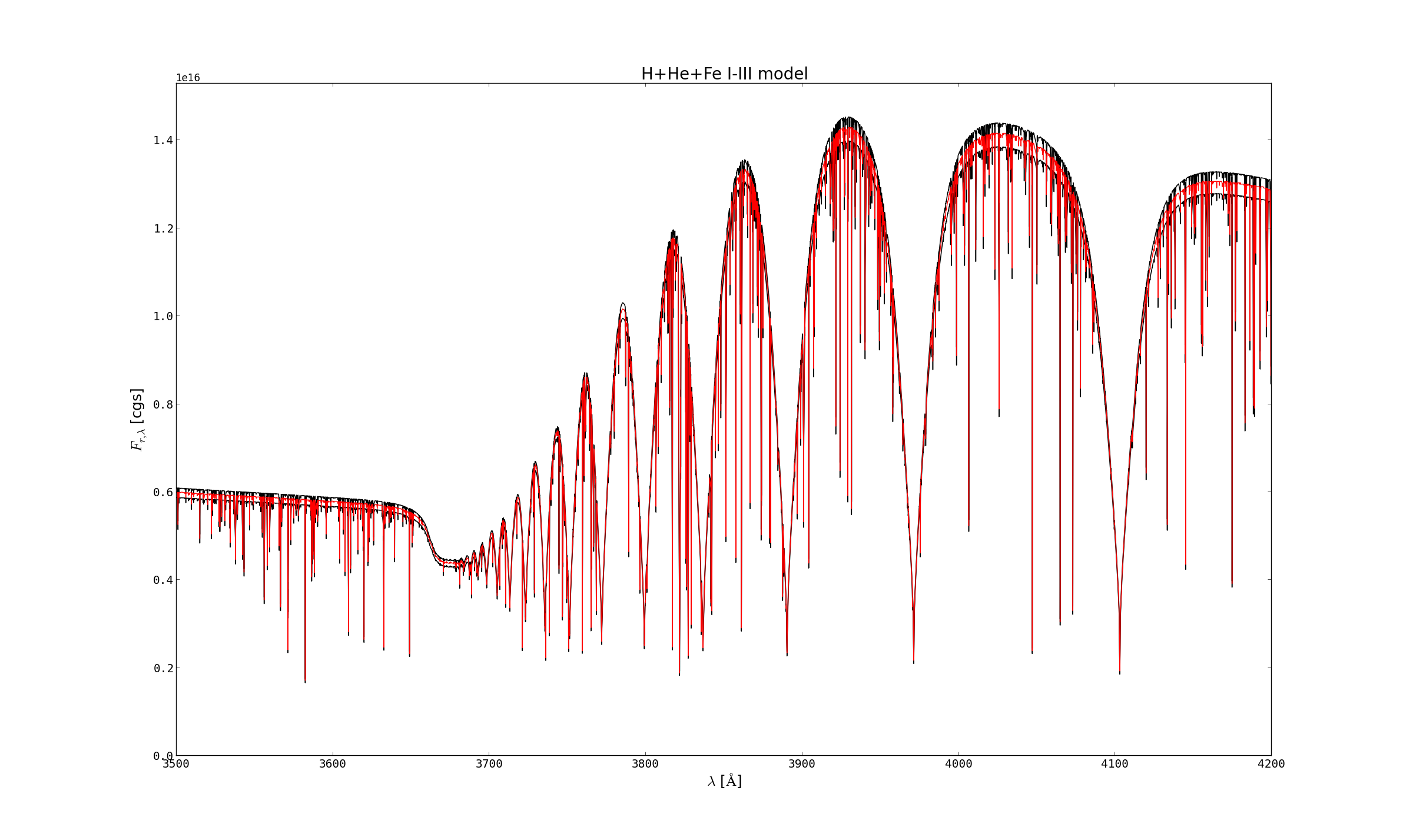}}
\caption{\label{fig:HHeFe:Balmer_HHeFe}
For our third test case with NLTE treatment of H~I, He~I--II,
and Fe~I--III,  the
spectral region near the Balmer jump is shown. 
The red lines show the results of the 1D NLTE calculation
whereas the black lines show the maximum and minimum of the radial component 
of the flux vector over all outermost voxels from the 3D calculation.
The spread in the 3D model is due to the finite numerical resolution,
see paper VI.
}
\end{figure*}

\begin{figure*}
\centering
\resizebox{\hsize}{!}{\includegraphics[angle=00]{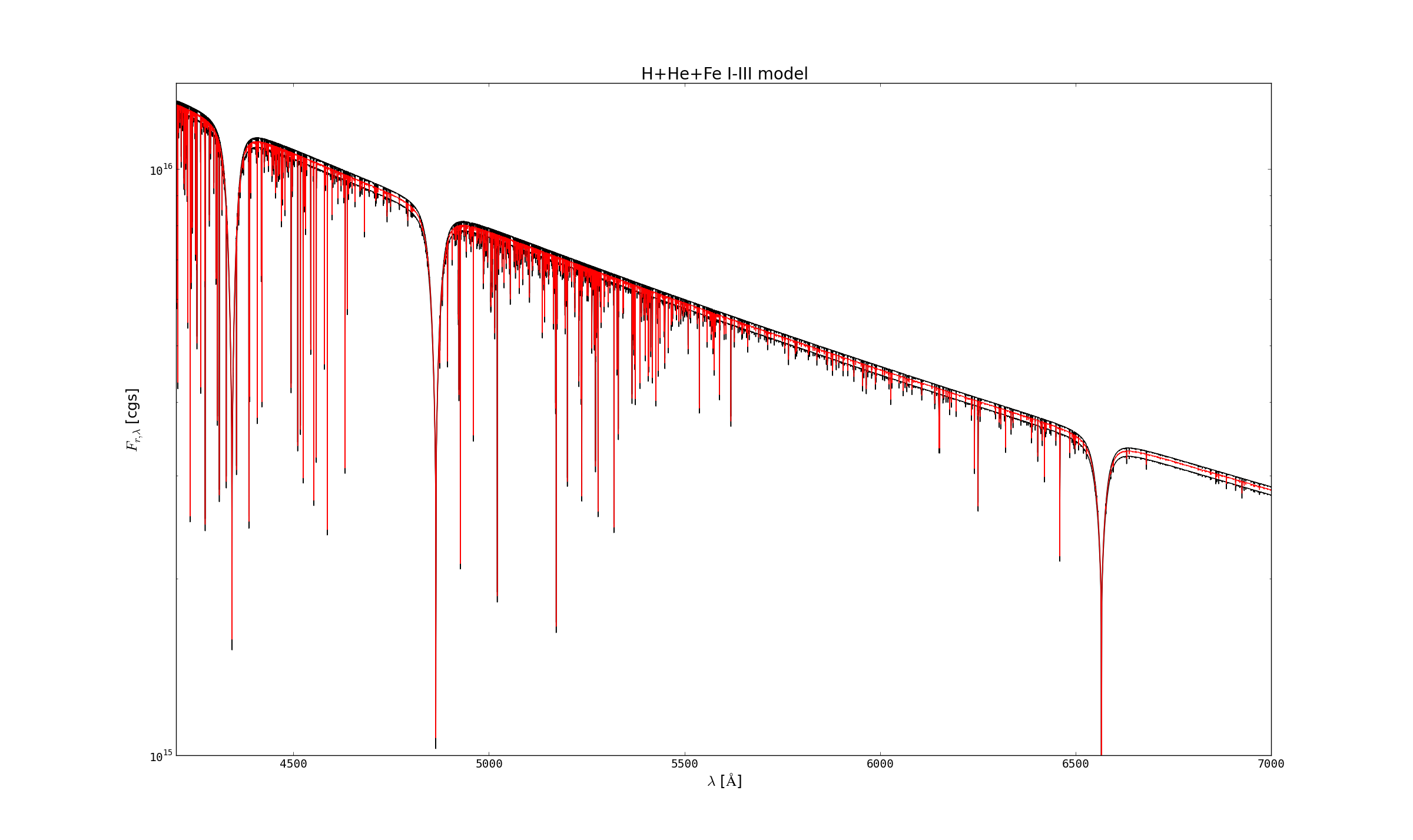}}
\caption{\label{fig:HHeFe:Balmer2_HHeFe}
For our third test case with NLTE treatment of H~I, He~I--II,
and Fe~I--III,  the
spectral region in the optical is shown. 
The red lines show the results of the 1D NLTE calculation
whereas the black lines show the maximum and minimum of the radial component 
of the flux vector over all outermost voxels from the 3D calculation.
The spread in the 3D model is due to the finite numerical resolution,
see paper VI.
}
\end{figure*}

\begin{figure*}
\centering
\resizebox{\hsize}{!}{\includegraphics[angle=00]{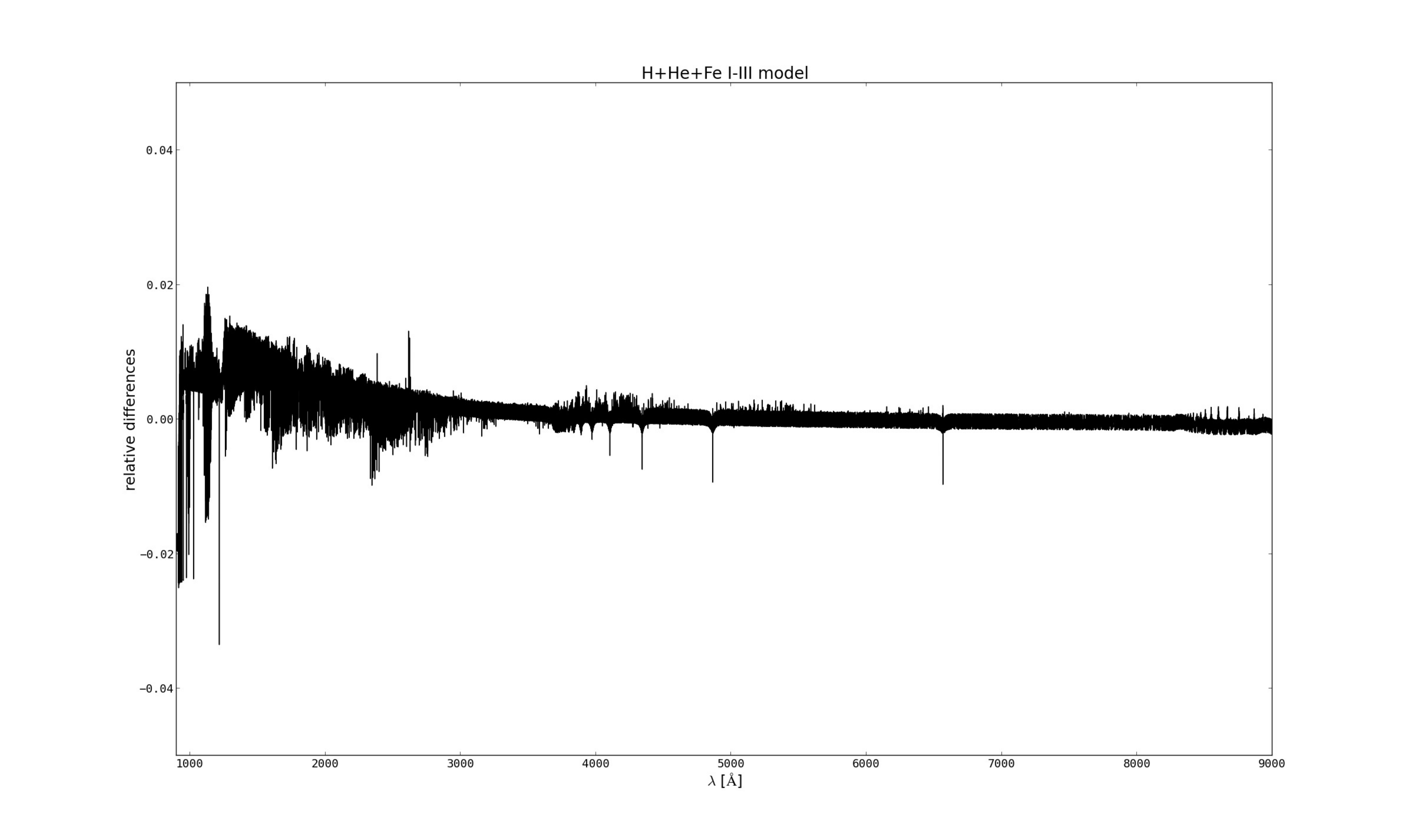}}
\caption{\label{fig:HHeFe:RelDiff}
Relative differences between the fluxes of the 1D comparison model and the
arithmetic average over all outermost voxels of the $r$ component of the flux vector of the 3D
calculation for the test case with NLTE treatment of H~I, He~I--II, and
Fe~I--III.
}
\end{figure*}

In order to test a  very large model (in present day terms), we repeated the
calculation for a setup with H, He and Fe I-III in NLTE. In this case, the Fe
model atoms are significantly larger than the model atoms used before.  With
the same setup for H and He as above, we have now Fe I (902 levels, 24395 lines
) Fe II (894 levels, 22453 lines) and Fe III (555 levels, 9867 lines), for a
total of 57232 individual lines and 2410 b-f transitions.  The number of levels
per Fe ion is so large, what we plot only every 10th level in
Fig.~\ref{fig:HHeFe:bi_2600} for clarity.  The agreement between 1D and 3D
models is again excellent, the differences are below the variances caused by
the resolution of the 3D model.  In Figs.~\ref{fig:HHeFe:Balmer_HHeFe} --
\ref{fig:HHeFe:Balmer2_HHeFe} we show two wavelength ranges comparing 1D (red
lines) and 3D spectra. The plot shows the maximum and minimum over all
outermost  voxels, the spread
is due to the finite numerical resolution, \citep[see][Paper VI]{3drt_paper6}.
The comparison  in these cases is also  very good.
 We show 
as an example in Fig.~\ref{fig:HHeFe:RelDiff} the relative differences between the 
1D comparison model and the arithmetic mean of the $r$ components of the 3D flux
vectors over all outermost voxels. With the exception of a few wavelengths, the 
differences are below 2\%, which is very good given the limits of the resolution in
solid angle.
The spectrum of
this model contains close  to 0.5 million wavelength points, the storage
requirements for the 3D spectral data (all voxels, complete information for
imaging including flux vectors) is about 2.6TB.

\section{Summary and Conclusions}

In this paper we  discussed a method to solve 3D multi-level radiative
transfer problems with detailed model atoms. The method is a direct extension
of the well-tested method we are using for 1D model atmosphere calculations.
We have implemented this NLTE/3D module for \phxT\ and discussed a small to very
large test calculations for code testing and validation. The results show that the 
method performs in 3D exactly as the 1D equivalent. The NLTE problem in 3D
poses significant demands on the computing resources, therefore we 
designed the module for distributed memory parallel processing (MPI) 
and to use domain decomposition methods to reduce the memory requirements
per process. With this, it is technically possible to even solve
3D NLTE problems for complex ions, e.g., the iron group, if large
supercomputers are used. 

\begin{acknowledgements}
 We thank the referee for providing really helpful comments and suggestions that
improved the original manuscript significantly.
This work was supported in part by DFG GrK 1351 and SFB 676, as well as NSF
grant AST-0707704.  The work has been supported in part by support for programs
HST-GO-12298.05-A, and HST-GO-122948.04-A  was provided by NASA through a grant
from the Space Telescope Science Institute, which is operated by the
Association of Universities for Research in Astronomy, Incorporated, under NASA
contract NAS5-26555.  The calculations presented here were performed partially
at the H\"ochstleistungs Rechenzentrum Nord (HLRN) and at the National Energy
Research Supercomputer Center (NERSC), which is supported by the Office of
Science of the U.S.  Department of Energy under Contract No. DE-AC03-76SF00098.
We acknowledge PRACE for awarding us access to resource JUQUEEN
based in Germany at the J\"ulich Supercomputing Centre (JSC).  We thank all these
institutions for a generous allocation of computer time.  The authors
gratefully acknowledge the Gauss Centre for Supercomputing (GCS) for providing
computing time through the John von Neumann Institute for Computing (NIC) on
the GCS share of the supercomputer JUQUEEN at J\"ulich Supercomputing Centre
(JSC). GCS is the alliance of the three national supercomputing centres HLRS
(Universit\"at Stuttgart), JSC (Forschungszentrum J\"ulich), and LRZ (Bayerische
Akademie der Wissenschaften), funded by the German Federal Ministry of
Education and Research (BMBF) and the German State Ministries for Research of
Baden-W\"urttemberg (MWK), Bayern (StMWFK) and Nordrhein-Westfalen (MIWF).
\end{acknowledgements}

\bibliography{rte_paper2,yeti,radtran,stars}

\clearpage

\end{document}